\documentclass[twocolumn,aps,prl,showpacs,amsmath,longbibliography,superscriptaddress,notitlepage]{revtex4-2}

\usepackage{amssymb}
\usepackage{mathrsfs}
\usepackage{graphicx}
\usepackage{float}
\usepackage[caption=false]{subfig}
\usepackage[pdftex,colorlinks=red]{hyperref}
\usepackage{titlesec}

\usepackage{hyperref}
\hypersetup{
    colorlinks=true,    
    linkcolor=blue,     
    citecolor=blue,     
    urlcolor=blue       
}
\usepackage{natbib}

\usepackage{amsmath}
\usepackage{epstopdf}
\usepackage[normalem]{ulem}
\usepackage{verbatim}
\usepackage{xcolor}
\usepackage{bm}

\def\red{\textcolor{black}}

\usepackage{titlesec}
\renewcommand\thesection{\Roman{section}}
\titleformat{\section}[block]{\filcenter\normalsize\bfseries}{\thesection.}{1em}{}
\renewcommand{\thesubsection}{\Alph{subsection}} 
\titleformat{\subsection}[block]{\bfseries\centering}{\thesubsection.}{1em}{} 

\begin{document}

\title{\red{Concurrent Skin-scale-free Localization} and Criticality under M\"obius Boundary Conditions in a Non-Hermitian Ladder}
\author{Shu Long}
\affiliation{Guangdong Provincial Key Laboratory of Quantum Metrology and Sensing $\&$ School of Physics and Astronomy, Sun Yat-sen University (Zhuhai Campus), Zhuhai 519082, China}\author{Linhu Li}\email{lilinhu@quantumsc.cn}
\affiliation{Quantum Science Center of Guangdong-Hong Kong-Macao Greater Bay Area, Shenzhen, China}

\begin{abstract}
Non-Hermitian systems possess exotic localization phenomena beyond their Hermitian counterparts, exhibiting massive accumulation of eigenstates at the system boundaries with different scaling behaviors. In this study, we investigate two weakly coupled non-Hermitian Hatano-Nelson chains under M\"obius boundary conditions (MBCs), and reveal the coexistence of two distinct localization behaviors for eigenstates. Namely, eigenstates exhibit non-Hermitian skin effect in one chain and scale-free localization in the other. Notably, the localization characteristics of eigenstates can exchange between the two chains depending on their eigenenergies. This phenomenon is found to emerge from the critical behaviors induced by the weak interchain coupling, which can even be enhanced by MBCs in comparison \red{to} the system under other boundary conditions. Our findings deepen the understanding of non-Hermitian localization and criticality, and offer new insights into engineering tunable edge-localized states in synthetic quantum systems.
\end{abstract}

\maketitle

\setcounter{secnumdepth}{1}

\section{INTRODUCTION}
Non-Hermitian Hamiltonians provide \red{a} simple but efficient description for non-equilibrium systems, such as open quantum systems and wave systems with gain or loss~\cite{ashida2020non,bender2007making,rotter2009non,persson2000observation,choi2010quasieigenstate,reiter2012effective,volya2003non}.
In non-Hermitian lattice systems, one fascinating feature is the presence of non-reciprocal hopping amplitudes, which leads to the non-Hermitian skin effect (NHSE) under open boundary conditions (OBCs)~\cite{yao2018edge,yokomizo2019non}. 
Systems with NHSE have loop-like spectra in the complex energy plane under periodic boundary conditions (PBCs), which shrink into curves when the boundary is opened, with all eigenstates localized at the boundaries ~\cite{hu2024topological,lin2023topological,okugawa2020second,yao2018non,lee2019anatomy,okuma2020topological,borgnia2020non,zhang2020correspondence}.
Experimentally, the NHSE has been observed in systems such \red{as} acoustics~\cite{zhang2021acoustic,zhang2021observation,xiong2024tracking}, photonic\red{s}~\cite{lin2024observation,wang2024effective}, cold atoms~\cite{cai2024non}, electric circuits~\cite{liu2021non,zhang2023electrical,zou2021observation,liu2023experimental,li2025observation}, and \red{mechanical systems}~\cite{li2024observation}.

Systems with NHSE have been shown to be extremely sensitive to perturbation~\cite{budich2020non,li2020critical}. 
For instance, Ref. \cite{li2020critical} reported the discovery of a critical NHSE in a weakly coupled non-Hermitian two-chain system, where eigenstates exhibit scale-free localization (SFL), with the localization length scaling proportionally to the system size. 
Furthermore, SFL has been identified both theoretically and experimentally in non-Hermitian systems in other scenarios, including a boundary impurity in one-dimensional (1D) non-Hermitian lattices~\cite{li2021impurity,wang2025observation}, local non-Hermiticity in Hermitian ones~\cite{guo2023accumulation,li2023scale,xie2024observation},
and long-range non-Hermitian couplings~\cite{ke2023topological}. 
The exotic SFL has further inspired investigation into non-Hermitian localizations with different scaling properties, such as the hybrid scale-free skin effect~\cite{fu2023hybrid}, the scale-tailored localization~\cite{guo2024scale}, and localization with a two-scale wavefunction~\cite{davies2025two}.

In this paper, we extend the investigation of non-Hermitian localization to a two-chain lattice with translational symmetry broken by the M\"obius boundary conditions (MBCs), as illustrated in Fig. \ref{fig:1}.
In the weakly coupled regime, we discover an intriguing localization feature with SFL on one chain and NHSE on the other, dubbed as the \red{concurrent skin-scale-free localization}.
We systematically analyze the system's spectral properties and eigenstate distributions, and unveil transitions between SFL and NHSE on different chains.
Furthermore, the system \red{exhibits} a real-complex transition of eigenenergies when turning on the weak interchain coupling, reflecting the critical NHSE that leads to the SFL.
In contrast to OBC systems,
critical behaviors here survive when the system is gapped by a strong on-site energy detuning between the two chains, indicating an enhancement of criticality by MBCs.

The remainder of this paper is organized as follows. In Sect. \ref{sec:model}, we introduce the basic information about the model and briefly discuss its properties under different boundary conditions. In Sect. \ref{sec:dua}, we unveil \red{concurrent skin-scale-free localization} by providing a detailed analysis of the energy spectra and localization characteristics of eigenstates for our model under various parameter regimes. In Sect. \ref{sec:cri}, we analyze the critical properties of the system. Finally, we summarize this paper in Sect. \ref{sec:con}.

\section{MODEL}\label{sec:model}

\begin{figure}[t] 
    \centering
    \includegraphics[scale=0.97]{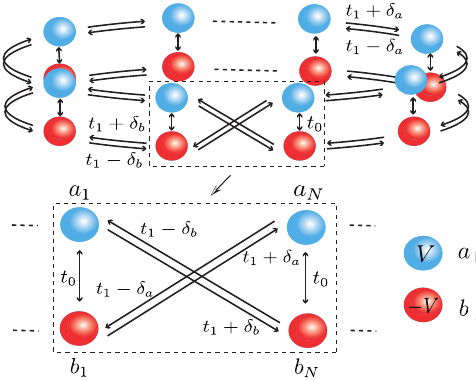}
    \caption{Schematic diagram of the model. Blue and red spheres denote chain $a$ and chain $b$, respectively, with on-site potentials $V$ and $-V$. The interchain coupling strength is denoted by $t_0$, while the non-reciprocal nearest-neighbor hopping amplitudes are given by $t_1  \pm \delta_s$ ($s=a,b$). The hopping amplitudes at the boundary are detailed in the lower panel. }
    \label{fig:1}
\end{figure}

We consider two coupled Hatano-Nelson (HN) chains~\cite{hatano1996localization} with MBCs, as schematically illustrated in Fig. \ref{fig:1}. The tight-binding Hamiltonian is
\begin{equation}
   \hat H = {{\hat H}_0} + {{\hat H}_{\rm BC}},
   \label{eq1}
\end{equation}
with the bulk term
\begin{equation}
  \begin{aligned}
   {{\hat H}_0}  =& \sum\limits_{x = 1}^{N - 1} {\sum\limits_{s=a,b} {({t_1} + {\delta _s}){{\hat s}^\dag }_{x + 1}{{\hat s}_x} + ({t_1} - {\delta _s}){{\hat s}^\dag }_x{{\hat s}_{x + 1}}} } \\
   &{\rm{ + }}\sum\limits_{x = 1}^N {[{t_0}(\hat a_x^\dag {{\hat b}_x} + \hat b_x^\dag {{\hat a}_x})}  + V(\hat a_x^\dag {{\hat a}_x} - \hat b_x^\dag {{\hat b}_x})],\\
     \end{aligned}  \label{eq:OBC}
\end{equation}
and the boundary term
\begin{equation}
  \begin{aligned}  
   {{\hat H}_{\rm BC}} =&\,\  ({t_1} - {\delta _a})\hat b_1^\dag {{\hat a}_N} + ({t_1} - {\delta _b})\hat a_1^\dag {{\hat b}_N}\\
   &+ ({t_1} + {\delta _a}){{\hat a}^\dag }_N{{\hat b}_1} + ({t_1} + {\delta _b}){{\hat b}^\dag }_N{{\hat a}_1}.
   \end{aligned}  
\end{equation}
Here, $N$ denotes the number of unit cells, $s=a,b$ \red{represents} the two chains, and ${{{\hat s}^\dag}_{x} }$ (${\hat s}_x$) is the creation (annihilation) operator on chain $s$ at the $x$-th unit cell. The term ${t_1} \pm {\delta _s}$ represents the non-reciprocal nearest-neighbor hopping amplitudes, while $t_0$ is the interchain coupling strength. An on-site energy detuning of $2V$ is applied between the two chains. Under MBCs, hopping amplitudes connect lattice sites $a_1$ and $b_N$ as well as $b_1$ and $a_N$. For simplicity, the hopping amplitude between $b_N$ and $a_1$ is set to match that of chain $b$, while the hopping amplitude between $a_N$ and $b_1$ is set to be the same as that of chain $a$, as shown in the lower panel in Fig. \ref{fig:1}.

\begin{figure}[t] 
    \centering
    \includegraphics[width=1\linewidth]{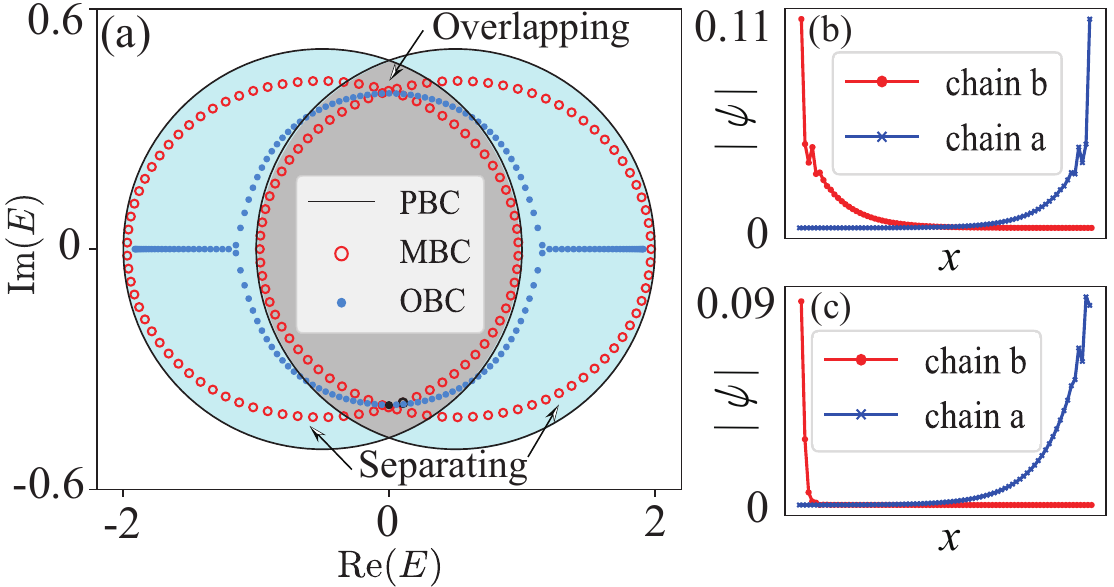}
    \caption{Energy spectra and representative eigenstates of the model. 
    (a) Energy spectra under PBCs (black lines), OBCs (blue dotes), and MBCs (red circles).
    (b) Eigenstate associated with the black dot in the OBC spectrum.
    (c) Eigenstate associated with the black circle in the MBC spectrum. Other parameters are $N= 80$, $t_1=0.75$, $\delta_a=0.25$, $\delta_b=-0.25$, $t_0=0.01$, and $V=0.5$. 
    }
    \label{fig:2}
\end{figure}

By taking PBCs for the bulk Hamiltonian $\hat{H}_0$ and transforming it into momentum space, we get
\begin{equation}
   {H_0}(k) = \sum\limits_k {\left. {\left( {\begin{array}{*{20}{c}}
   {\hat{c}_{{k_a}}^\dag }&{\hat{c}_{\red{{k_b}}}^\dag }
   \end{array}} \right.} \right)} \left( {\begin{array}{*{20}{c}}
   {{f_a}(k)}&{{t_0}}\\
   {{t_0}}&{{f_b}(k)}
   \end{array}} \right)\left( {\begin{array}{*{20}{c}}
   {{\hat{c}_{{k_a}}}}\\
   {{\hat{c}_{{k_b}}}}
   \end{array}} \right),
\end{equation}
where ${f_a}(k) = ({t_1} - {\delta _a}){e^{{\rm i}k}} + ({t_1} + {\delta _a}){e^{ - {\rm i}k}} + V$, ${f_b}(k) = ({t_1} - {\delta _b}){e^{{\rm i}k}} + ({t_1} + {\delta _b}){e^{ - {\rm i}k}} - V$. Solving the characteristic equation $\det[H_0(k)-E(k)I]=0$, we can obtain the PBC energy spectrum, given by
\begin{equation}
   \begin{aligned}
E(k) = &\,\,2{t_1}\cos k - {\rm i}({\delta _a} + {\delta _b})\sin k\\
 &\pm \sqrt { - \frac{1}{2}{{({\delta _a} - {\delta _b})}^2} + t_0^2 + {V^2} + \frac{1}{2}({\delta _a} - {\delta _b})[h(k)]},  
\end{aligned}
\end{equation}
with $h(k) = ({\delta _a} - {\delta _b})\cos 2k + 4{\rm i} V\sin k$. The PBC spectrum, represented by black lines in Fig.~\ref{fig:2}(a), forms two partially-overlapped loops in the complex plane. For positive $V$ and small $t_0$, the right loop can be attributed to the energy spectrum of chain $a$ and the left loop can be attributed to chain $b$.
For clarity of later discussion, we label an overlapping region and two separating regions in Fig.~\ref{fig:2}(a), according to the relation between the two loops of PBC spectrum.

For ${{\hat H}_{\rm BC}}=0$, the system is under OBCs, and its Hamiltonian, within the framework of the generalized Brillouin zone (GBZ)~\cite{yao2018edge,yokomizo2019non,zhang2020correspondence}, is given by  \begin{equation}
  {H_{\rm{OBC}}}(\beta ) = \left( {\begin{array}{*{20}{c}}
  {{g_a}(\beta )}&{{t_0}}\\
  {{t_0}}&{{g_b}(\beta )}
  \end{array}} \right), 
\end{equation}
with ${g_a}(\beta ) = ({t_1} - {\delta _a})\beta  + ({t_1} + {\delta _a}){\beta ^{ - 1}} + V$ and ${g_a}(\beta ) = ({t_1} - {\delta _b})\beta  + ({t_1} + {\delta _b}){\beta ^{ - 1}} - V$. In this case, the system exhibits characteristics of the critical NHSE at weak $t_0$, as extensively analyzed in Refs. \cite{li2020critical,yokomizo2021scaling}. The corresponding OBC energy spectrum is represented by blue dots in Fig. \ref{fig:2}(a), which consists \red{of} a point-gapped loop spectrum centered at $E=0$ that falls mostly in (or close to) the overlapping region, and two branches of line spectra with positive and negative real energies in the separating regions, respectively. An OBC eigenstate associated with the black dot in Fig. \ref{fig:2}(a) is shown in Fig. \ref{fig:2}(b). As illustrated, the states on the two chains are symmetric. Unlike the conventional non-Hermitian skin effect, it has been shown in Ref.~\cite{li2020critical} that the amplitude of such an eigenstate \red{follows} $\left| \psi_{a/b, x} \right| \sim {e^{ - x/N}}$, where $\psi_{a/b,x}$ denotes the wave amplitude on chain $a$ or $b$ at the $x$-th site. Consequently, the spatial distribution of eigenstates is independent from the system size $N$ after normalization, giving rise to the phenomenon known as SFL. 

Owing to the pronounced sensitivity of non-Hermitian systems to boundary conditions, it is reasonable to expect that MBCs introduce unique properties to the system. We \red{perform} preliminary numerical calculations to obtain the energy spectrum, represented by red circles in Fig.~\ref{fig:2}(a). Our results show that the energy spectrum under MBCs closely resembles that under PBCs, consisting two loops in the complex plane. 
On the other hand, the eigenstates are localized at the boundaries with asymmetric distributions on the two ends of the system.
A representative MBC eigenstate, corresponding to the black circle in Fig.~\ref{fig:2}(a), is shown in Fig.~\ref{fig:2}(c). 
\red{It is seen that localization on the right end of chain $a$ is similar to the SFL in Fig.~\ref{fig:2}(b); yet much stronger localization is observed on the left end of chain $b$, which hints the NHSE. The distinct scaling behaviors of these two types of localization in our system will will be discussed in detail in Sec.~\ref{sec:dua}.}

\section{\red{CONCURRENT SKIN-SCALE-FREE LOCALIZATION}}\label{sec:dua}

The distinct localization profiles along the two chains in Fig.~\ref{fig:2}(c) suggest the coexistence of NHSE and SFL in single eigenstates under MBCs, dubbed as the \red{concurrent skin-scale-free localization}.
To justify this phenomenon, in this section we provide a detailed analysis of the spectral and localization properties of the system, in parameter regions with different non-reciprocal pumping directions of the two chains.
Note that we have fixed $t_0\ll 1$ for the system to stay in the critical regime; 
otherwise the \red{concurrent skin-scale-free localization} will be destroyed with all eigenstates becoming delocalized (see Appendix \ref{A1}).

\subsection{A. Opposite Non-Reciprocal Hopping}

\begin{figure}[t]
    \centering
    \includegraphics[scale=0.42]{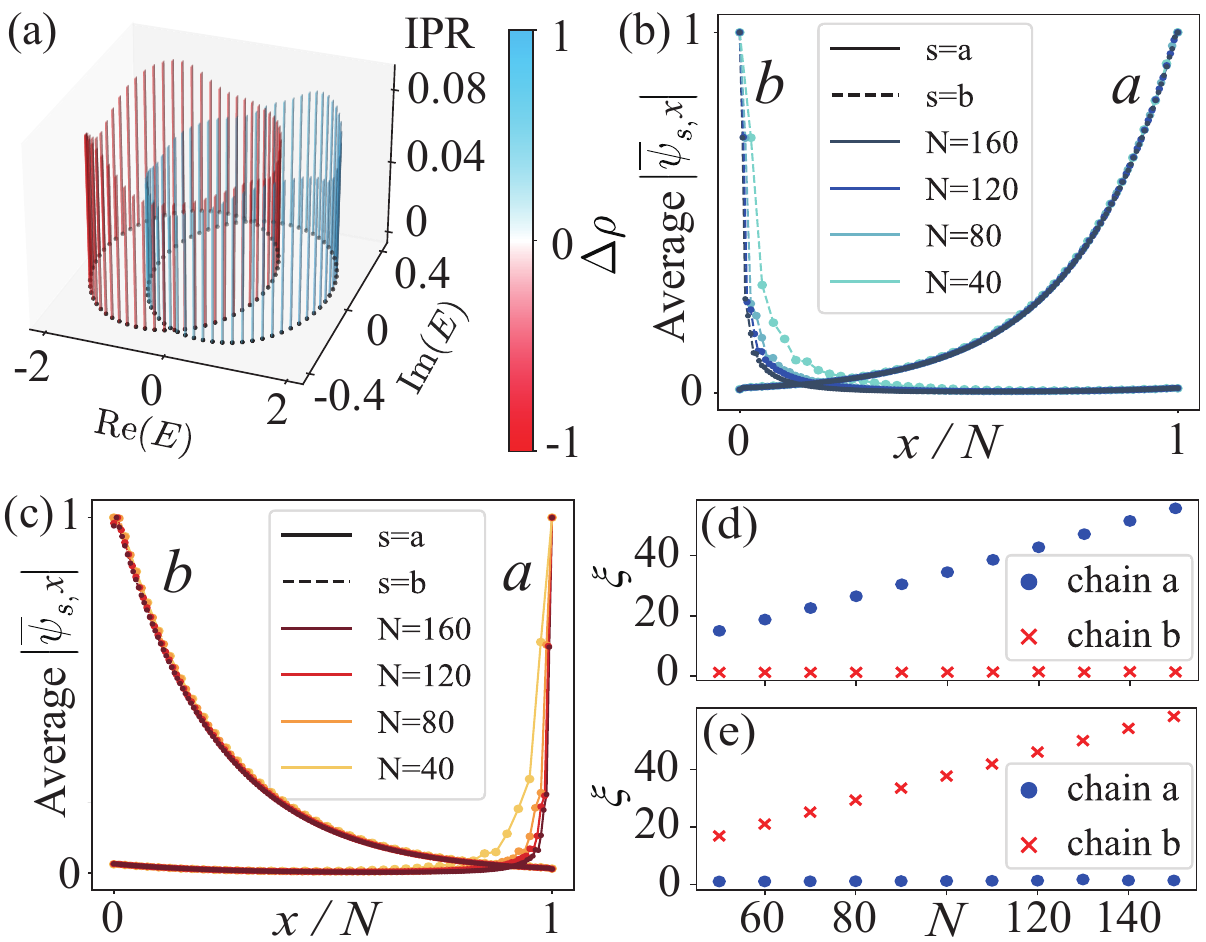} 
    \caption{\red{concurrent skin-scale-free localization} with opposite non-reciprocities on the two chains. (a) Energy spectrum for $N=60$. Height of vertical lines represents the values of IPR for corresponding eigenstates, and colors indicate values of $\Delta\rho$, their single-chain polarization.
    (b) and (c) depict the average distribution of all eigenstates with positive and negative $\Delta\rho$ in (a), respectively.
    Distribution along chain $a$ is plotted with solid lines,
    and that along chain $b$ \red{is} ploted with dashed lines.
    Dark and light colors represent systems with different sizes $N$, which are normalized to $1$ for comparison. The eigenstate magnitudes are rescaled such that the maximum value of the average distribution is set to 1.
    SFL is characterized by the identical localization profiles, while NHSE is characterized by the localization profiles varying with $N$.
    (d) and (e) single-chain localization length $\xi$ of the average distribution in (b) and (c), respectively. 
    Other parameters are $t_1=0.75$, $\delta _a =0.25$, $\delta _b =-0.25$, $t_0=0.01$, and $V=0.5$.
    }
    \label{fig:3}
\end{figure}

We first consider the case where non-reciprocities on the two chains have the same magnitude, but are opposite to each other, namely with $\delta_a=-\delta_b$.
The complex spectrum under MBCs \red{is} displayed in Fig. \ref{fig:3}(a), where each eigenenergy is marked \red{by} the inverse participation ratio (IPR)
\begin{equation}
{\rm{IP}}{{\rm{R}}} = {\sum_{s=a,b}\sum\limits_{x = 1}^N {\left| {{\psi _{s,x}}} \right|} ^4}
\end{equation}
and the single-chain density polarization
\begin{equation}
\Delta \rho {\rm{ = }}{\rho _a} - {\rho _b} = {\sum\limits_{x = 1}^N {\left| {{\psi _{a,x}}} \right|} ^{\rm{2}}} - {\sum\limits_{x = 1}^N {\left| {{\psi _{b,x}}} \right|} ^{\rm{2}}}
\end{equation}
of the corresponding eigenstate $|\psi\rangle$.
All eigenstates are seen to possess a certain degree of localization, indicated by their ${\rm IPR}\gg 1/2N$.
On the other hand, with a weak interchain \red{coupling} $t_0$, the spectrum forms two partially overlapped loops, reflecting a pseudo-decoupled scenario where each loop can be adiabatically connected to the PBC spectrum of a single chain.
Consistently, eigenstates are found to possess $|\Delta\rho|\gg0$, indicating that they distribute mainly on one of the two chains.
\red{For clarity, we define the chain where an eigenstate mostly occupies as 
the dominant chain of the state.}

Despite the strong single-chain polarization, since the two chains are connected end to end through MBCs, their maximum wave amplitudes may take approximately the same value for some eigenstates, as shown by the example in Fig.~\ref{fig:2}(c). For such states, the large density on the dominant chain can be attributed to the slow density decay of scale-free localization into the bulk, in contrast to the rapid one of skin localization.
Consequently, the small density on the \red{less-occupied} chain cannot be simply ignored, as its local maximum value can approach that of the dominant chain.
In addition, we note that the small but non-vanishing density on the \red{less-occupied} chain 
also distinguishes the pseudo-decoupled system from the strongly-decoupled case with gapped spectrum, as will be further discussed in Sec. \ref{sec:changingV} B.

To unveil the localization properties of these states, we defined a single-chain distribution with its maximum amplitude normalized to 1,
\begin{equation}
|{\bar \psi }_{{\rm s},x}| = \frac{|\psi _{s,x}|}{{\rm max}_x[\left| {\psi _{s,x}} \right|]},
\end{equation}
and present its average for states with $\Delta\rho >0$ and $\Delta\rho <0$ in Fig. \ref{fig:3}(b) and (c), respectively.
Specifically, while all eigenstates show exponential localization along both chains,
we find that those with $\Delta\rho>0$ possess a size-independent profile along chain $a$, indicating the SFL; 
and a stronger localization along chain $b$ as the system size $N$ increases, indicating the NHSE.
Similarly, a reverse phenomenon is observed for eigenstates with $\Delta\rho<0$, where SFL occurs along chain $b$ and NHSE occurs along chain $a$.

To further verify this \red{concurrent skin-scale-free localization},
we extract the localization length $\xi$ from
\begin{equation}
  \psi_{s,x} \sim A{e^{ - \frac{{\left| {x - {x_0}} \right|}}\xi }}
  \label{eq12}
\end{equation}
through numerical fitting, where $A$ is a normalization factor and $x_0$ is the center of the chain. 
In Fig. \ref{fig:3}(d), we show the results of $\xi$ obtained
from the average distribution of eigenstates with $\Delta\rho>0$. The blue dots represent the localization length $\xi$ along chain $a$, which exhibits a linear dependence on the system size, $\xi\sim \alpha N$, justifying the SFL.
On the other hand, the red dots indicate a size-independent $\xi$ along chain $b$, reflecting the feature of NHSE. 
Finally, in Fig. \ref{fig:3}(e) for states with $\Delta\rho<0$, the localization characteristics exchange between the two \red{chains}, \red {consistent with} our observation in Fig. \ref{fig:3}(c).

\subsection{B. Identical Non-Reciprocal Hopping}

\begin{figure}[ht]
    \centering
    \includegraphics[scale=0.40]{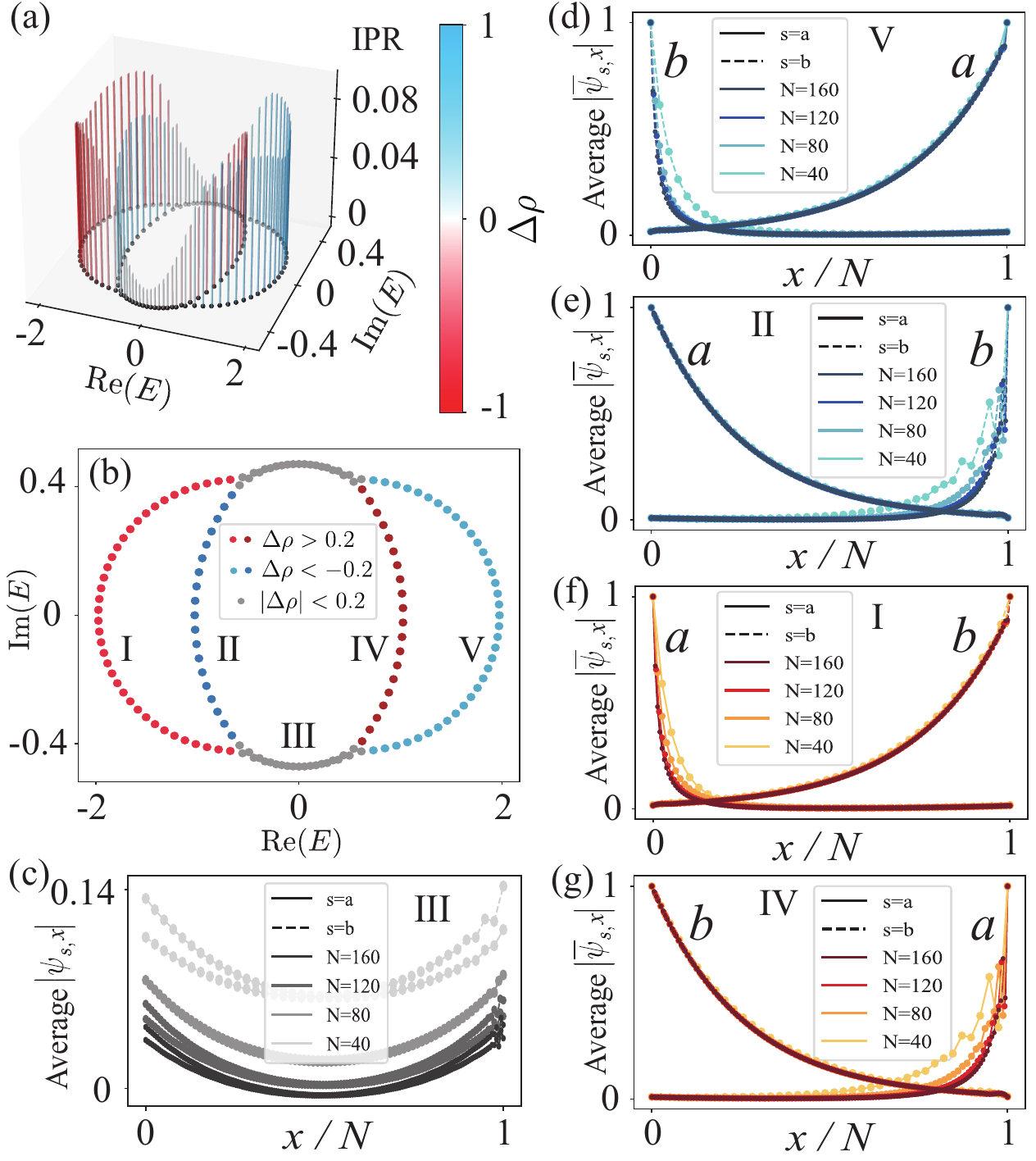} 
    \caption{
    \red{Concurrent skin-scale-free localization} with identical non-reciprocities on the two chains. 
    (a) Energy spectrum for $N=60$. Height of vertical lines \red{represent} the values of IPR for corresponding eigenstates, and colors indicate values of $\Delta\rho$, their single-chain polarization. 
    (b) Energy spectrum for $N=80$. Eigenstates with different $\Delta\rho$ values are marked by colors and roman numbers I to V.
    \red{(c) shows the average distribution of eigenstates of regions III in (b). Distribution along chain $a$ is plotted with solid lines,
    and that along chain $b$ is plotted with dashed lines.
    Dark and light colors represent systems with different sizes $N$, which are normalized to $1$ for comparison.}
    (d), (e), (f) and (g) depict the average distribution of eigenstates of regions V, II, I, and IV in (b), respectively. \red{The methods used to indicate colors and line types are the same as those in (c).}
    The eigenstate magnitudes are rescaled such that the maximum value of the average distribution is set to 1.
    SFL is characterized by the identical localization profiles, while NHSE is characterized by the localization profiles varying with $N$. Notably, eigenstates in regions II and IV show reversed left-SFL on the chain they primary occupy, against the non-reciprocal direction of the system.   
    Other parameters are $t_1=0.75$, $\delta _a =\delta _b =0.25$, $t_0=0.01$, and $V=0.5$.
    }
    \label{fig:4}
\end{figure}

Next we examine another typical case with $\delta_a=\delta_b$, namely, the two chains have identical non-reciprocities.
It does not support critical behaviors or SFL under OBCs, as the two chains share the same GBZ.
However, MBCs inhomogeneously \red{connect} the two \red{ends} of the system, acting as a local impurity that may also induce SFL.

In Fig. \ref{fig:4}(a), we display the MBC eigenenergies marked by IPR and $\Delta\rho$ of each eigenstate.
It is seen that the energy spectrum now forms a unified structure of a loop in the center with two "earrings" in contrast to the two overlapping loops when $\delta_a=-\delta_b$.
Such an inseparable structure indicates that the pseudo-decoupled picture is no longer valid for the whole system; indeed, some eigenstates with ${\rm Re}(E)\approx0$ possess small values of IPR and vanishing $\Delta\rho$, corresponding to extended states distribute evenly on the two chains \red{[see Fig.~\ref{fig:4}(c)]}.
Meanwhile,
most eigenstates still possess large IPR and $|\Delta\rho|$, showing strong localization and single-chain polarization as in the previous case.

To have a comprehensive understanding of the localization,
we divide the eigenenergies into five groups according to their real energies and degrees of single-chain polarization (reflected by the density polarization $\Delta\rho$), as shown in
Fig. \ref{fig:4}(b).
The average distributions for eigenstates for groups with non-ignorable single-chain polarization ($|\Delta\rho|>0.2$ in our numerical simulation) are plotted in Fig. \ref{fig:4}(d) to (g), respectively.
The existence of \red{concurrent skin-scale-free localization} for these eigenstates is characterized by the co-existence of SFL and NHSE on different chains.
Specifically, as shown in Fig.~\ref{fig:4}(f) and (d), groups I and V show SFL align with the non-reciprocal direction (toward the right in our example), with eigenstates mostly occupying chain $b$ and $a$, respectively.
Physically, this is because the onsite energy $V$ shifts the two chains away from each other in real energy, and these two groups mostly fall in the separating regions of the two loops [see Fig.~\ref{fig:2}(a)].
Thus, the corresponding localization properties can still be understood with the pseudo-decoupled picture. \red{It is worth noting that, in this case, the skin states emerge on the side opposite to what would be expected from the system's non-reciprocity. We attribute the origin of these skin states to the decay of the wavefunction in the less-occupied chain from the edge where the dominant chain is localized, as the two chains are connected end-to-end by MBCs.
Thus the skin localization is independent of the non-reciprocal hopping in the less-occupied chain. 
Importantly, we argue that these are not merely boundary-induced defect states. Instead, they are rather manifestations of a skin effect governed by both bulk and boundary contributions, because (i) their number scales porpotionally to the system size; and (ii) the localization vanishes in biorthogonal distribution (see Appendix \ref{A3}), indicating its origin of global non-Hermiticity instead of local defect.}

More intriguingly, groups II and IV exhibit a reversed SFL against the non-reciprocal direction~\cite{li2021impurity}, as shown in Fig.~\ref{fig:4}(e) and (g).
Note that normal and reversed SFL may occur in a single-chain model with fixed non-reciprocity, but different impurity strengths~\cite{li2021impurity}.
In turn, the coexistence of normal and reversed SFL in our model may be attributed to the different degrees of coupling between the two chains.
That is, the total effective non-reciprocity may \red{vary} for eigenstates in the pseudo-decoupled (separating) and coupled (overlapping) regions, making them react to the same boundary impurity (MBCs) differently.

\subsection{C. General Non-Reciprocal Hopping Amplitudes}

\begin{figure}[t] 
    \centering
    \includegraphics[scale=0.31]{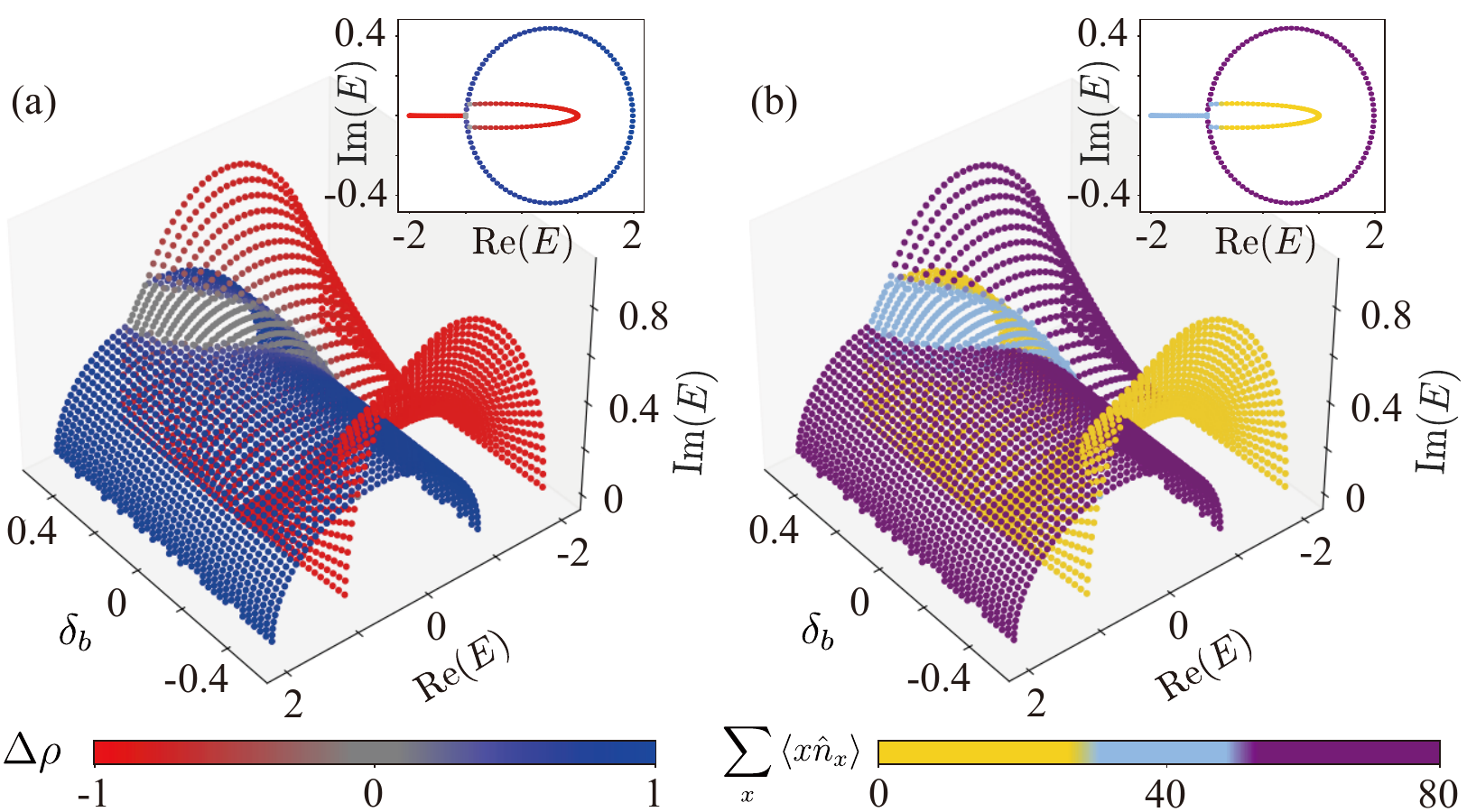}
    \caption{\red{Concurrent skin-scale-free localization} with different non-reciprocities on the two chains.
    (a) and (b) display the energy spectra under different $\delta_b$ values. In (a), eigenenergies are marked by colors according to the \red{single-chain} polarization $\Delta\rho$, whereas in (b), they are marked by the state's average position $\sum_x\langle x\hat{n}_x\rangle$.
    The insets show the spectra at $\delta_b=0$, annotated with their respective quantities from the main figures.
    Other parameters are $N=80$, $t_1=0.75$, $\delta _a =0.25$, $t_0=0.01$, and $V=0.5$.   
    }
    \label{fig:5}
\end{figure}

After examining the two extreme cases, we now investigate how 
localization properties change with varying the non-reciprocal nearest-neighbor hopping amplitudes. 
In Fig. \ref{fig:5}(a), we plot the eigenenergies (with only ${\rm Im}(E)\ge0$ for a clearer view) with different $\delta_b$.
Most eigenstates are seen to \red{be} polarized either on chain $a$ (blue) or $b$ (red), while some unpolarized states (gray) emerge when $\delta_b>0$.

In Fig. \ref{fig:5}(b), we display the same eigenenergies marked by colors according to the average position of their eigenstates,
\begin{equation}
 \sum_{x}\langle x\hat{n}_x \rangle=\sum_{x}\langle x(\hat{a}_x^\dagger\hat{a}_x+\hat{b}_x^\dagger\hat{b}_x) \rangle.   
\end{equation}
The average position reflects the direction of SFL for the skin-scale-free states, 
since it manifests on the dominant chain.
We find that when the non-reciprocal direction of chain $b$ (\red{indicated} by the sign of $\delta_b$) reverses, 
SFL also reverses its direction for eigenstates with ${\rm Re}(E)<0$, despite that some of them are polarized on chain $a$ [blue colors with ${\rm Re}(E)<0$ in Fig. \ref{fig:5}(a)].
In comparison, eigenstates with ${\rm Re}(E)>0$ keep their SFL direction unchanged for positive and negative $\delta_b$, even for those polarized on chain $b$ [red colors with ${\rm Re}(E)>0$ in Fig. \ref{fig:5}(a)].

Finally, we note another transition occurring at $\delta_b=0$, where chain $b$ \red{becomes} Hermitian. As shown in the insets of Fig. \ref{fig:5}, a branch of eigenenergies with ${\rm Re}[E]<0$ are real when $\delta_b=0$. 
Their corresponding eigenstates have $\Delta\rho\approx -1$ and $\sum_x\langle x \hat{n}_x\rangle\approx N/2$,
corresponding to extended bulk states mostly occupying the Hermitian chain $b$ with a negative potential $-V$.

\section{CRITICALITY UNDER MBCS}\label{sec:cri}
As shown in previous discussion and Appendix \ref{A1}, concurrent skin-scale-free states emerge only with a weak interchain coupling $t_0$, indicating that they are a class of critical states analogous to the scale-free states in the same model under OBCs~\cite{li2020critical}.
In this section, we analyze the criticality of the MBC system based on its spectral characteristics and unveil that the criticality can even be enhanced by the MBCs.

\subsection{A. Real-complex Transition of Eigenenergies}\label{sec:complex_energy}

\begin{figure}[t] 
    \centering
    \includegraphics[width=1.0\linewidth]{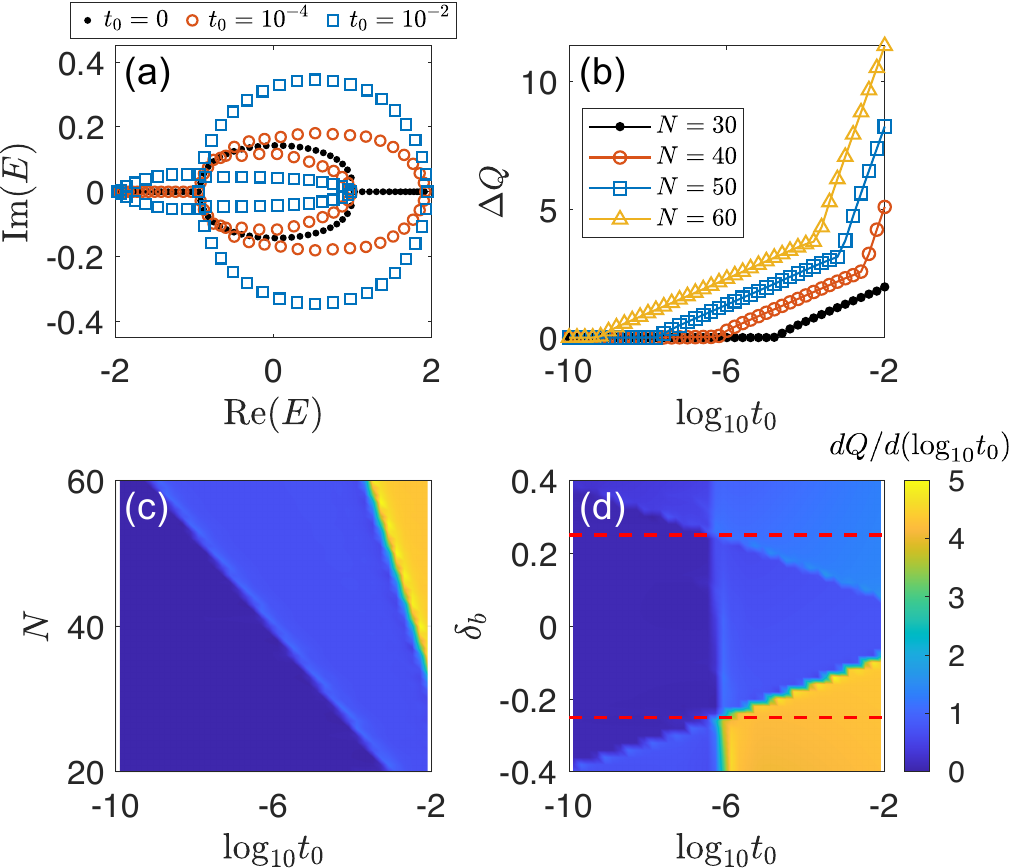}
    \caption{ 
    (a) Energy spectra of the model under MBCs with different interchain coupling strength $t_0$.
    The two branches of real eigenenergies at $t_0=0$ acquire \red{imaginary} values at different nonzero $t_0$.
    (b) The increase in the total imaginary magnitude of eigenenergies caused by $t_0$, $\Delta Q$, as a function of $t_0$ for different the system with sizes $N$.
    The sudden changes in the slope indicate the critical values of $t_0$ where the two branches of real eigenergies in (a) become complex.
    (c) Derivative of the total imaginary magnitude of eigenenergies versus $N$ and $t_0$. 
    The critical values of $t_0$ are reflected by the jumps of the derivative, and tend to be zero in the thermodynamic limit $N\rightarrow\infty$.
    (d) Derivative of the total imaginary magnitude of eigenenergies versus $\delta_b$ and $t_0$.
    Boundaries between different plateaux cross each other when  $|\delta_b|=|\delta_a|$ (red dashed lines), where the two chains have identical parameters (possibly with opposite non-reciprocity), and the system has a single critical value of $t_0$ for the real-complex transition of eigenenergies.
    Other parameters are $V=0.5$, $t_1=0.75$, $\delta_a=0.25$, $\delta_b=-0.1$, $t_0=0.01$, $N=40$, unless \red{otherwise} specified in the figure.}
    \label{fig:6}
\end{figure}

In Ref.~\cite{li2020critical}, it has been shown that the two-chain model under OBCs, described by the Hamiltonian $H_0$ in Eq.~\eqref{eq:OBC}, 
\red{exhibits} the critical NHSE characterized by a real-complex transition of eigenenerges~\footnote{Or more generally, a line-loop transition of the spectrum.} induced by a weak $t_0$.
A similar transition also occurs under MBCs, as shown by the eigenenergies with different values of $t_0$ in Fig. \ref{fig:6}(a).
That is, the system under MBCs processes two branches of real eigenenergies when $t_0=0$, which acquire imaginary values at different $0<t_0\ll1$. 

To characterize this transition in the spectrum,
we introduce the total imaginary magnitude of eigenenergies, $Q=\sum |{\rm Im}(E)|$, and its increment $\Delta Q$ caused by $t_0$, defined as
\begin{eqnarray}
\Delta Q=Q(t_0)-Q(t_0=0).
\end{eqnarray}
As shown in Fig. \ref{fig:6}(b), $\Delta Q$ remains constant for small $t_0$, and begins to increase when $t_0$ exceeds a critical value, reflecting the real-complex transition of the eigenenergies.
The slope is seen to increase twice with $t_0$, indicating the different critical $t_0$ for the two branches of real eigenenergies at $t_0=0$. 
This is because, due to the real-energy detuning $V$, the two branches of real eigenenergies are mainly contributed by different single chains.
Thus, the critical value of $t_0$ is expected to be different for them when the two chains possess distinct strengths of non-reciprocity (namely, $|\delta_a|\neq|\delta_b|$ in our model).

Another important observation from Fig.~\ref{fig:6}(b) is that the critical value of $t_0$ decreases as the size of the system increases,
analogous to the critical NHSE under OBCs~\cite{li2020critical}.
To demonstrate the transition of eigenenergies more clearly, we show the derivative of $Q$ with respect to $t_0$ in Fig.~\ref{fig:6}(c) for different system's \red{sizes} $N$. The jumps of the derivative \red{indicate} the critical value of $t_0$, which tends to zero when $N\rightarrow \infty$.
Note that when $t_0=0$, the system becomes a junction of two 1D non-Hermitian chains that presents NHSE~\cite{deng2019non}. 
Therefore, we can conclude that in the thermodynamic limit, an infinitesimal interchain coupling can induce a significant change in the spectrum accompanied by a transition from skin states to concurrent skin-scale-free states, which represent a type of critical NHSE.

A crucial feature of the critical NHSE under MBCs is that it occurs for eigenstates in the separating regions of PBC spectrum [see Fig.\ref{fig:2}(a)], in contrast to that under OBCs which occurs mostly in (or close to) the overlapping region~\cite{li2020critical}.
More precisely,
the critical transition here emerges in each single chain when it is subjected to the influence of the other chain through the interplay between MBCs and interchain coupling.
One evidence is the observation of a two-step real-complex transition of the spectrum, where eigenenergies in the two separating regions acquire imaginary values at two independent values of $t_0$.
As shown in Fig. \ref{fig:6}(d), only one of the two critical $t_0$ varies with $\delta_b$, 
corresponding to the transition in chain $b$. Meanwhile, the other critical $t_0$ remains unchanged since parameters in chain $a$ remain constants ($t_0\approx 10^{-6}$ in our example). 
Notably, the two critical values of $t_0$ coincide when $\delta_b=\delta_a$ (red dashed lines in the figure), where the two chains become identical.
However, the real-complex transition still occurs, unlike the critical NHSE under OBCs which requires two non-identical chains~\cite{li2020critical}.

\subsection{B. Criticality under a Strong On-site Energy Detuning}\label{sec:changingV}

\begin{figure}[t] 
    \centering
    \includegraphics[scale=0.49]{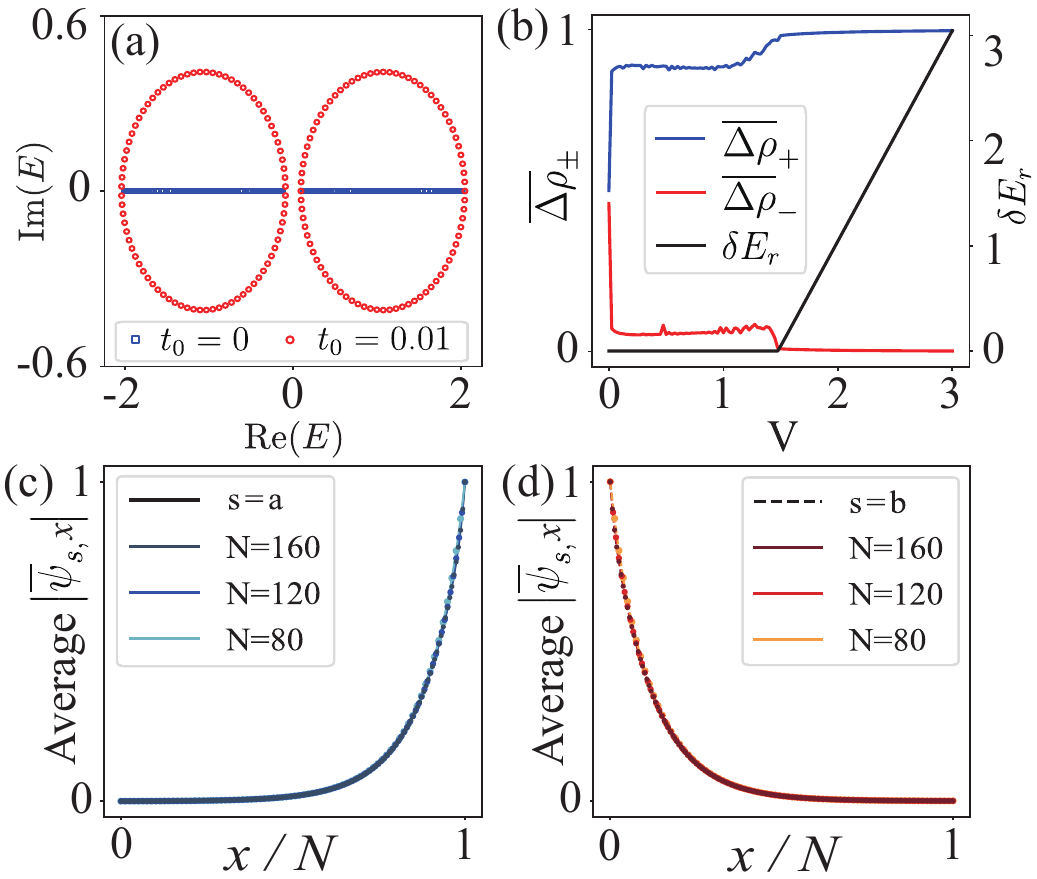}
    \caption{
    (a) Energy spectrum for $V = 1.6$. The red dots represent the spectrum with a small $t_0$, while the blue dots correspond to the case of $t_0 = 0$.
    (b) The blue and red lines illustrate the average density $\overline{\Delta\rho}_\pm$, respectively, as a function of $V$. The black line represents the variation of the energy gap $\delta E_r$ along the Re($E$) axis as a function of $V$. The system size is $N = 60$. 
    (c) and (d) show the average modulus of all eigenstates for $V = 30$ corresponding to $\Delta\rho > 0$ and $\Delta\rho < 0$, respectively.
    Other than the parameter values indicated in the figure, the remaining parameters are chosen as follows: $t_1 = 0.75$, $\delta_a = 0.15$, $\delta_b = -0.15$, and $t_0 = 0.01$. 
    }
    \label{fig:7}
\end{figure}

To further demonstrate and analyze the criticality of each single chain, we \red{increase} the energy detuning $V$ between the two chains to separate them in eigenenergies. 
As shown in Fig. \ref{fig:7}(a), the spectrum now forms two bands gapped in real energy when $V > \sqrt{2t_1^2 - t_0^2}$.
In this case, when $t_0=0$, the MBC eigenenergies become real as under OBCs, suggesting that the two chains are fully decoupled and exhibit the normal NHSE. 
However, such a system still \red{acquires} complex eigenenergies when introducing a weak $t_0$, indicating the criticality which otherwise vanishes in the OBC system with gapped spectrum (see Appendix \ref{A2}).

On the other hand, 
despite both possessing critical behaviors indicated by their energy characteristics,
a gapped system shows different eigenstate distribution \red{compared} with previous cases without a gap in real energy. To see this,
we define the average density polarization for the two bands,
\begin{eqnarray}
\overline{\Delta\rho}_+=\sum_{\Delta\rho>0}\Delta\rho/N_+,~~~~\overline{\Delta\rho}_-=\sum_{\Delta\rho<0}\Delta\rho/N_-,
\end{eqnarray}
with $N_+$ ($N_-$) the number of eigenstates with positive (negative) $\Delta\rho$, which are the same as those with positive (negative) real eigenenergies when the two bands are gapped.
In Fig. \ref{fig:7}(b), we plot $\overline{\Delta\rho}_\pm$ as a function of $V$ and compare it with the variation of real-energy gap $\delta E_r$.
It is observed that as $V$ increases, $\overline{\Delta\rho}_\pm$ jumps to $\pm1$ when the real-energy gap opens ($\red{\delta E_r}>0$), 
indicating a strongly-decoupled scenario where eigenstates occupy almost completely in one of the two chains, and the distribution on the other chain becomes negligible.
This is in contrast to the pseudo-decoupled scenario in Sec.~\ref{sec:dua} with small but non-negligible $\Delta\rho$, corresponding to the region with $\delta E_r=0$ in Fig. \ref{fig:7}(b).

Finally, we demonstrate the average distribution of eigenstates with positive and negative real energies Figs. \ref{fig:7}(c) and \ref{fig:7}(d), respectively.
Note we show only the distribution along the dominant chain, as density on the other chain is negligible in this case.
We find that eigenstates still exhibit SFL even for a large $V$ that induces a gapped spectrum.
Such a feature and the loop-like spectrum in Fig. \ref{fig:7}(a) reflect the enhanced criticality associated with $t_0$, and
highlight the significant impact of MBCs in this strongly-decoupled scenario,
distinguishing it from the system under PBCs (with extended states due to translational symmetry) and OBCs (with real spectrum and NHSE) where the weak $t_0$ can be safely ignored when the two bands are gapped.
\red{Physically, this is because the system with $t_0=0$ are fully-decoupled under OBCs or PBCs. Thus, from a perturbation point of view, a weak $t_0$ coupling the two chains only affects the spectrum through higher-order processes. However, MBCs mix the two chains and thus leading to non-vanishing first-order perturbation from $t_0$ (see Appendix \ref{A4}).}

\red{Alternatively, the enhanced criticality can also be understood from the topological braiding of MBCs. Explicitly, MBCs braid the two chains and connect them into a longer one, where $t_0$ becomes highly non-local long-range coupling. Such nonlocality may also lead to SFL~\cite{wang2023scaling}, which is a representative feature of the criticality in our model.}

\section{CONCLUSIONS AND DISCUSSION}\label{sec:con}
In this paper, we unveil a class of localization phenomena characterized by a \red{concurrent skin-scale-free localization} in non-Hermitian two-chain systems under MBCs. 
The origin of this behavior can be summarized as follows: a weak interchain coupling ($t_0$) drives the system into a critical regime featuring SFL in the dominant chain, while the MBCs connect the two chains end to end, ensuring a non-vanishing density on the \red{less-occupied} chain that manifests NHSE. The localization directions of NHSE and SFL for each eigenstate are opposite to each other, and can be tuned as desired by adjusting the non-reciprocal hopping directions in each chain. 
Furthermore, depending the parameters, the system can be classified into several cases with distinct spectral and eigenstate characteristics:
\begin{itemize}
\item{the pseudo-decoupled scenario under MBCs, with \red{concurrent skin-scale-free localization}, non-negligible density in the \red{less-occupied} chain, and a gapless spectrum;}
\item{the strongly-decoupled scenario under MBCs, with SFL in the dominant chain, negligible density in the \red{less-occupied} chain, and a gapped spectrum; }
\item{the fully-decoupled scenario under OBCs, with NHSE along each chain, and a gapped spectrum. }
\end{itemize}
These rich localization behaviors provide a versatile platform for engineering and manipulating exotic edge-localized states in synthetic lattice systems. \red{A possible physical implementation of the MBCs is through electric circuits. Similar circuit models have already been designed in Ref.~\cite{li2020critical}, and different boundary conditions (including MBCs) can be implemented as desired by connecting the two ends of the circuit lattice properly.
}

Moreover, the critical behavior observed under MBCs exhibits key distinctions from that under other boundary conditions. In particular, the criticality of each chain is found to be independent, despite being coupled via $t_0$ and the MBCs. That is, the threshold value of $t_0$ required to induce SFL and trigger a real-complex transition of eigenenergies depends solely on the parameters of the corresponding chain. Interestingly, we also find that criticality can be enhanced by MBCs. That is, SFL states with real-complex spectral transitions persist even under strong energy detuning between the two chains—a regime in which the spectrum is gapped and critical behavior is otherwise suppressed under OBCs.
These distinct critical and localization behaviors
not only provide a versatile platform for engineering and manipulating exotic edge-localized states in synthetic lattice systems,
but also underscore the rich and boundary-sensitive nature of non-Hermitian criticality, extending beyond the conventional paradigms established under other boundary conditions.

\section{ACKNOWLEDGMENT}
L. L. acknowledges helpful discussion with Sen Mu.
This work is supported by 
the National Natural Science Foundation of China (Grant No. 12474159).

\appendix
\renewcommand{\thesection}{\Alph{section}} 
\titleformat{\section}[block] 
  {\normalsize\bfseries\filcenter} 
  {APPENDIX \thesection} 
  {1em} 
  {} 

\renewcommand{\theequation}{\thesection\arabic{equation}} 

\section{MBC SYSTEM WITH STRONG COUPLING}\label{A1}

\begin{figure}[h] 
    \centering
    \includegraphics[scale=0.42]{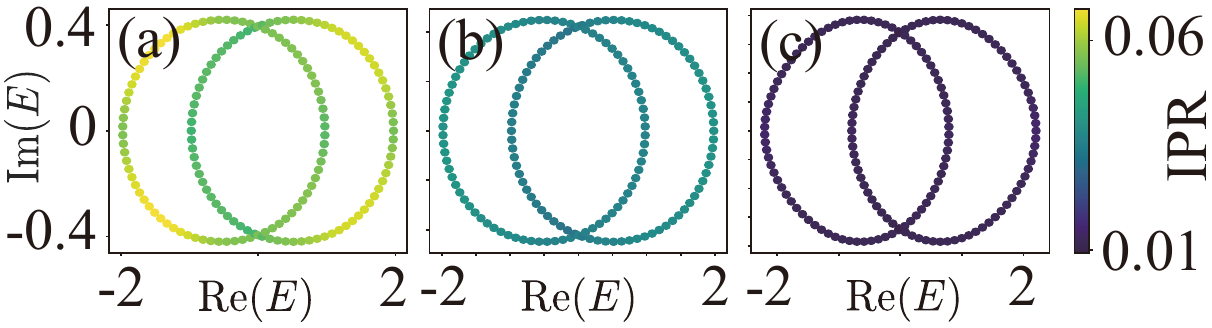}
    \caption{Energy spectra for different $t_0$ values. (a), (b), and (c) correspond to $t_0=0.01$, $0.05$, and $0.5$, respectively. The color scale represents different IPR values of eigenstates at each point. Other parameters are chosen as $N=80$, $t_1=0.75$, $\delta _a =-\delta_b=0.25$, and $V=0.5$.}
    \label{fig:A1}
\end{figure}

The energy spectra under MBCs for different values of $t_0$ are shown in Fig. \ref{fig:A1}. As $t_0$ increases, the shape of the spectrum undergoes slight changes. However, when $t_0$ deviates from the critical regime, all states become delocalized (as shown by the vanishing IPR), indicating the breakdown of the \red{concurrent skin-scale-free localization}.

\section{\red{SINGLE EIGENSTATE IN DIFFERENT BASIS}}\label{A3}

\begin{figure}[t] 
    \centering
    \includegraphics[scale=0.61]{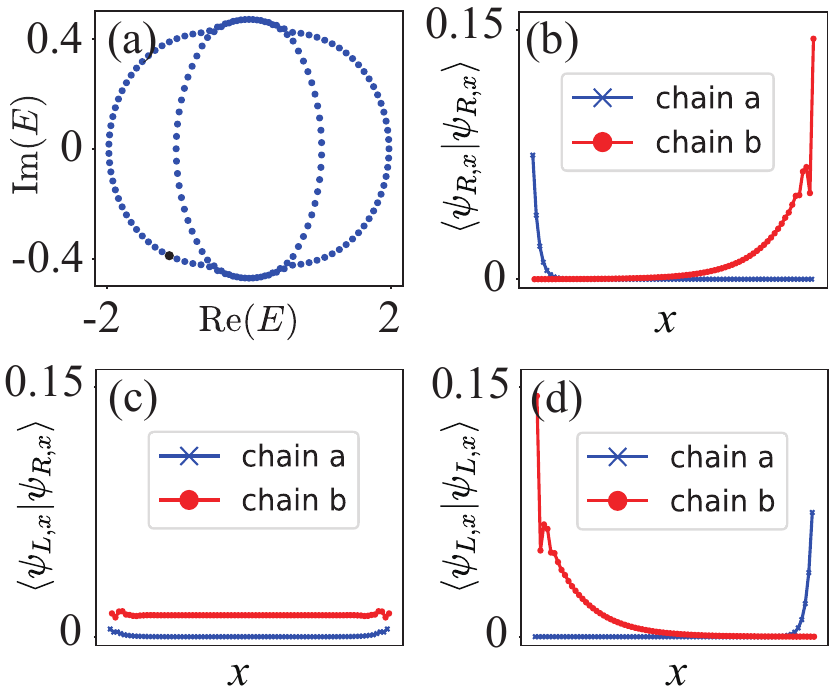}
    \caption{\red{Energy spectrum and single eigenstates associated
    with the black dot in different basis. (a) Energy spectrum. (b), (c), and (d) are $\langle\psi_{R,x}|\psi_{R,x}\rangle$, $\langle\psi_{L,x}|\psi_{R,x}\rangle$, and $\langle\psi_{L,x}|\psi_{L,x}\rangle$, respectively.  Other parameters are chosen as $N=80$, $t_1=0.75$, $\delta_a =\delta_b=0.25$, and $t_0=0.01$.}}
    \label{fig:A3}
\end{figure}

\red{
To verify whether the localized states discussed in the main text arise solely as defect states induced by the MBCs, we perform calculations of the states in the biorthogonal basis.
Generally, the eigenvalue equations of a non-Hermitian Hamiltonian,
\begin{equation} \hat{H}|\psi_{R,x}\rangle=E|\psi_{R,x}\rangle,\,\,\,\hat{H}^\dagger|\psi_{L,x}\rangle=E^*|\psi_{L,x}\rangle,
\end{equation}
show that the right and left eigenvectors $|\psi_{R,x}\rangle$ and $|\psi_{L,x}\rangle$ differ from each other. Fig.~\ref{fig:A3} presents the results of inner products computed using different combinations of left and right eigenvectors. Panel (c) shows the results in the biorthogonal basis, which show that the distribution is roughly uniform throughout the system and primarily occupies one chain. Although MBCs break the translational symmetry and induce slight fluctuations in the biorthogonal distribution near the boundaries, these effects are much weaker than the localization observed when considering only the right or left eigenstates [see panels (b) and (d)]. Therefore, we conclude that the localized states discussed in the main text do not arise solely as defect states induced by the boundaries.
}

\section{OBC SYSTEM WITH GAPPED SPECTRUM}\label{A2}

\begin{figure}[h] 
    \centering
    \includegraphics[scale=0.52]{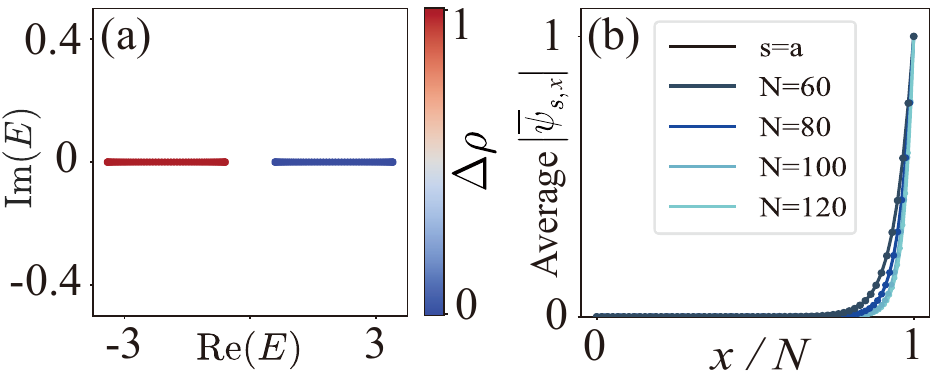}
    \caption{(a) The energy spectrum for $V=2$ ($N=80$) under OBCs, where different colors represent different $\Delta \rho$ values. (b) Rescaled eigenstates on chain $a$,  corresponding to the ${\rm Re}(E)>0$ band. Different colored lines indicate different $N$. Other parameters are chosen as $N=80$, $t_1=0.75$, $\delta _a =-\delta_b=0.25$, and $t_0=0.01$.}
    \label{fig:A2}
\end{figure}

Fig. \ref{fig:A2}(a) shows the energy spectrum under OBCs for $V = 2 > \sqrt{2t_1^2 - t_0^2}$, where all eigenvalues are real. Fig. \ref{fig:A2}(b) presents the average $|\overline{\psi}_{s,x}|$ of the band with $\mathrm{Re}(E) > 0$, indicating the absence of SFL in this case. Since the states in each band are almost entirely distributed on a single chain [see Fig. \ref{fig:A2}(a)] and the antisymmetric structure of the two chains, only the states on one chain from one band are shown.

{\section{\red{CRITICALITY ENHANCED BY MBCS}}\label{A4}
\red{Here we provide a perturbation treatment to explain how MBCs enhance criticality to a certain extent. Note that in the following we have taken a much simplified approximation of the unperturbed eigenstates, which does not give accurate results, but is enough for demonstrating the physical origin of our observations.}

\red{In our study, the criticality refers to the sensitivity of the system to a weak $t_0$. Thus, we may assume that $t_0$ is much weaker than the effect of MBCs. Therefore, we take MBCs as a perturbation to the system under OBCs, and $t_0$ as a higher-order perturbation to the overall system.
The explicit forms of these two terms are given by
\begin{align}
H_{t_0}=\sum_{x=1}^N t_0(\hat{a}_x^\dagger\hat{b}_x+\hat{b}_x^\dagger\hat{a}_x),
\end{align}
and
\begin{equation}
  \begin{aligned}  
   {{\hat H}_{\rm BC}} =&\,\  ({t_1} - {\delta _a})\hat b_1^\dag {{\hat a}_N} + ({t_1} - {\delta _b})\hat a_1^\dag {{\hat b}_N}\\
   &+ ({t_1} + {\delta _a}){{\hat a}^\dag }_N{{\hat b}_1} + ({t_1} + {\delta _b}){{\hat b}^\dag }_N{{\hat a}_1}.
   \end{aligned}  
\end{equation}
The strengths of these two terms are determined by $t_0$ and $t_1\pm\delta_s$, respectively.
In addition, we note that in the main text, the enhanced criticality is observed when the two chains are separated in energy by the on-site potential $V$. Thus, here we consider the scenario with $V\gg t_1\pm\delta_s \gg t_0$.}

\red{
First, the OBC system with $t_0=0$ is two separated Hatano-Nelson models. 
Expressing the eigenstates as
\begin{align}
\Psi=(\psi_{a,1},\psi_{a,2},...,\psi_{a,N},\psi_{b,1},\psi_{b,2},...,\psi_{b,N})^T,
\end{align}
The OBC eigenstates on chain $a$ and $b$ generally take the forms of~\cite{yao2018edge,okuma2020topological,lee2019anatomy}
\begin{align}
|\Psi_a^{\rm OBC}\rangle=C_a(r_a,r_a^2,...,r_a^N,0,0,...0)^T,\\
|\Psi_b^{\rm OBC}\rangle=C_b(0,0,...0,r_b,r_b^2,...,r_b^N)^T,
\end{align}
with $r_{s}=\sqrt{(t_1+\delta_{s})/(t_1-\delta_{s})}$ and $C_s$ the normalization factors.
Note that here we have taken approximate forms describing only the exponential decaying resultant from NHSE, and ignored the phase factor and oscillation from the Bloch wavefunction.
Without introducing MBCs, the first-order perturbation of $H_{t_0}$ is
\begin{align}
\langle \Psi_s^{\rm OBC}|H_{t_0}|\Psi_s^{\rm OBC}\rangle=0,
\end{align}
as each eigenstate distributes only on a single chain, and $H_{t_0}$ only contains coupling between the two chains.}

\red{
Next, we take MBCs as a perturbation to the OBC system.
The first-order perturbation to the eigenstates is given by
\begin{align}
|\Psi_a^{(1)}\rangle=&\sum_{\Phi_a^{\rm OBC}\neq\Psi_a^{\rm OBC}}\frac{\langle\Phi_a^{\rm OBC}|H_{\rm BC}|\Psi_a^{\rm OBC}\rangle}{E_{\Psi_a^{\rm OBC}}-E_{\Phi_a^{\rm OBC}}}|\Phi_a^{\rm OBC}\rangle\nonumber\\
&+\sum_{\Phi_b^{\rm OBC}}\frac{\langle\Phi_b^{\rm OBC}|H_{\rm BC}|\Psi_a^{\rm OBC}\rangle}{E_{\Psi_a^{\rm OBC}}-E_{\Phi_b^{\rm OBC}}}|\Phi_b^{\rm OBC}\rangle\nonumber\\
=&\frac{C_aC_bN}{V} \bigg[ (t_1-\delta_a)r_a^Nr_b+(t_1+\delta_b)r_ar_b^N \bigg] |\Psi_b^{\rm OBC}\rangle,
\end{align}
\begin{align}
|\Psi_b^{(1)}\rangle=&\sum_{\Phi_b^{\rm OBC}\neq\Psi_b^{\rm OBC}}\frac{\langle\Phi_b^{\rm OBC}|H_{\rm BC}|\Psi_b^{\rm OBC}\rangle}{E_{\Psi_b^{\rm OBC}}-E_{\Phi_b^{\rm OBC}}}|\Phi_b^{\rm OBC}\rangle\nonumber\\
&+\sum_{\Phi_a^{\rm OBC}}\frac{\langle\Phi_a^{\rm OBC}|H_{\rm BC}|\Psi_b^{\rm OBC}\rangle}{E_{\Psi_b^{\rm OBC}}-E_{\Phi_a^{\rm OBC}}}|\Phi_a^{\rm OBC}\rangle\nonumber\\
=&-\frac{C_aC_bN}{V} \bigg[ (t_1+\delta_a)r_a^Nr_b+(t_1-\delta_b)r_ar_b^N \bigg] |\Psi_a^{\rm OBC}\rangle.
\end{align}
The first term in each of these two equations vanishes, as $H_{\rm BC}$ also only contains coupling between the two chains.
In addition, we have further assumed  the $E_{\Psi_a^{\rm OBC}}=V$ and $E_{\Psi_b^{\rm OBC}}=-V$, as $V$ is the dominant energy scale in our consideration.
Note that despite the coefficient $1/V$, the amplitudes of these terms may not vanish for a large enough system size $N$.}

\red{
Thus, the eigenstates under MBCs, up to first-order approximation, are given by
\begin{align}
|\Psi_a^{\rm MBC}\rangle=|\Psi_a^{\rm OBC}\rangle+|\Psi_a^{(1)}\rangle,\\
|\Psi_b^{\rm MBC}\rangle=|\Psi_b^{\rm OBC}\rangle+|\Psi_b^{(1)}\rangle.
\end{align}
and we can already see the MBCs-induced mixture of the two chains for these eigenstates.
Thus, the first-order perturbation of $t_0$ to the eigenenergies under MBCs now takes non-vanishing value,
\begin{align}
\langle \Psi_s^{\rm MBC}|H_{t_0}|\Psi_s^{\rm MBC}\rangle\neq 0.
\end{align}
In other words, a system under MBCs is more sensitive to $t_0$ compared to that under OBCs, indicating an enhanced criticality.}

\bibliography{ref}

\end{document}